\RequirePackage[left]{lineno}
\documentclass[aps,prl,nofootinbib,twocolumn,showpacs,preprintnumbers,amsmath,amssymb,refmerge,superscriptaddress]{revtex4}
\usepackage{color}
\usepackage{graphicx}
\usepackage{dcolumn}
\usepackage{longtable}
\usepackage[percent]{overpic}

\def\babar  {\mbox{\slshape B\kern-0.1em{\smaller A}\kern-0.1em B\kern-0.1em{\smaller A\kern-0.2em R}}}

\newcommand{\bs}{\begin{sideways}}
\newcommand{\es}{\end{sideways}}
\newcommand{\bc}{\begin{center}}
\newcommand{\ec}{\end{center}}

\makeatletter
\newcommand*{\rom}[1]{\expandafter\@slowromancap\romannumeral #1@} 
\makeatother

\begin{document}

\preprint{\vbox{
                 \hbox{\bf Belle Preprint 2014-20}
                 \hbox{\bf KEK Preprint 2014-36}    
                 \hbox{}
}}

\title{Search for the dark photon and the dark Higgs boson at Belle}


\noaffiliation
\affiliation{University of the Basque Country UPV/EHU, 48080 Bilbao}
\affiliation{University of Bonn, 53115 Bonn}
\affiliation{Budker Institute of Nuclear Physics SB RAS and Novosibirsk State University, Novosibirsk 630090}
\affiliation{Faculty of Mathematics and Physics, Charles University, 121 16 Prague}
\affiliation{University of Cincinnati, Cincinnati, Ohio 45221}
\affiliation{Deutsches Elektronen--Synchrotron, 22607 Hamburg}
\affiliation{Department of Physics, Fu Jen Catholic University, Taipei 24205}
\affiliation{Justus-Liebig-Universit\"at Gie\ss{}en, 35392 Gie\ss{}en}
\affiliation{Gifu University, Gifu 501-1193}
\affiliation{The Graduate University for Advanced Studies, Hayama 240-0193}
\affiliation{Gyeongsang National University, Chinju 660-701}
\affiliation{Hanyang University, Seoul 133-791}
\affiliation{University of Hawaii, Honolulu, Hawaii 96822}
\affiliation{High Energy Accelerator Research Organization (KEK), Tsukuba 305-0801}
\affiliation{IKERBASQUE, Basque Foundation for Science, 48011 Bilbao}
\affiliation{Indian Institute of Technology Guwahati, Assam 781039}
\affiliation{Indian Institute of Technology Madras, Chennai 600036}
\affiliation{Institute of High Energy Physics, Chinese Academy of Sciences, Beijing 100049}
\affiliation{Institute of High Energy Physics, Vienna 1050}
\affiliation{Institute of Mathematical Sciences, Chennai 600113}
\affiliation{INFN - Sezione di Torino, 10125 Torino}
\affiliation{Institute for Theoretical and Experimental Physics, Moscow 117218}
\affiliation{J. Stefan Institute, 1000 Ljubljana}
\affiliation{Kanagawa University, Yokohama 221-8686}
\affiliation{Department of Physics, Faculty of Science, King Abdulaziz University, Jeddah 21589}
\affiliation{Korea Institute of Science and Technology Information, Daejeon 305-806}
\affiliation{Korea University, Seoul 136-713}
\affiliation{Kyungpook National University, Daegu 702-701}
\affiliation{\'Ecole Polytechnique F\'ed\'erale de Lausanne (EPFL), Lausanne 1015}
\affiliation{Faculty of Mathematics and Physics, University of Ljubljana, 1000 Ljubljana}
\affiliation{University of Maribor, 2000 Maribor}
\affiliation{Max-Planck-Institut f\"ur Physik, 80805 M\"unchen}
\affiliation{School of Physics, University of Melbourne, Victoria 3010}
\affiliation{Moscow Physical Engineering Institute, Moscow 115409}
\affiliation{Moscow Institute of Physics and Technology, Moscow Region 141700}
\affiliation{Graduate School of Science, Nagoya University, Nagoya 464-8602}
\affiliation{Kobayashi-Maskawa Institute, Nagoya University, Nagoya 464-8602}
\affiliation{Nara Women's University, Nara 630-8506}
\affiliation{National United University, Miao Li 36003}
\affiliation{Department of Physics, National Taiwan University, Taipei 10617}
\affiliation{H. Niewodniczanski Institute of Nuclear Physics, Krakow 31-342}
\affiliation{Niigata University, Niigata 950-2181}
\affiliation{Osaka City University, Osaka 558-8585}
\affiliation{Pacific Northwest National Laboratory, Richland, Washington 99352}
\affiliation{Peking University, Beijing 100871}
\affiliation{University of Pittsburgh, Pittsburgh, Pennsylvania 15260}
\affiliation{Soongsil University, Seoul 156-743}
\affiliation{Sungkyunkwan University, Suwon 440-746}
\affiliation{School of Physics, University of Sydney, NSW 2006}
\affiliation{Department of Physics, Faculty of Science, University of Tabuk, Tabuk 71451}
\affiliation{Tata Institute of Fundamental Research, Mumbai 400005}
\affiliation{Excellence Cluster Universe, Technische Universit\"at M\"unchen, 85748 Garching}
\affiliation{Toho University, Funabashi 274-8510}
\affiliation{Tohoku University, Sendai 980-8578}
\affiliation{Department of Physics, University of Tokyo, Tokyo 113-0033}
\affiliation{Tokyo Institute of Technology, Tokyo 152-8550}
\affiliation{University of Torino, 10124 Torino}
\affiliation{CNP, Virginia Polytechnic Institute and State University, Blacksburg, Virginia 24061}
\affiliation{Wayne State University, Detroit, Michigan 48202}
\affiliation{Yamagata University, Yamagata 990-8560}
\affiliation{Yonsei University, Seoul 120-749}
  
  \author{I.~Jaegle}\affiliation{University of Hawaii, Honolulu, Hawaii 96822} 
  \author{I.~Adachi}\affiliation{High Energy Accelerator Research Organization (KEK), Tsukuba 305-0801}\affiliation{The Graduate University for Advanced Studies, Hayama 240-0193} 
  \author{H.~Aihara}\affiliation{Department of Physics, University of Tokyo, Tokyo 113-0033} 
  \author{S.~Al~Said}\affiliation{Department of Physics, Faculty of Science, University of Tabuk, Tabuk 71451}\affiliation{Department of Physics, Faculty of Science, King Abdulaziz University, Jeddah 21589} 
  \author{D.~M.~Asner}\affiliation{Pacific Northwest National Laboratory, Richland, Washington 99352} 
  \author{T.~Aushev}\affiliation{Moscow Institute of Physics and Technology, Moscow Region 141700}\affiliation{Institute for Theoretical and Experimental Physics, Moscow 117218} 
  \author{R.~Ayad}\affiliation{Department of Physics, Faculty of Science, University of Tabuk, Tabuk 71451} 
  \author{A.~M.~Bakich}\affiliation{School of Physics, University of Sydney, NSW 2006} 
  \author{V.~Bansal}\affiliation{Pacific Northwest National Laboratory, Richland, Washington 99352} 
\author{M.~Barrett}\affiliation{University of Hawaii, Honolulu, Hawaii 96822} 
% ...
\collaboration{The Belle Collaboration}



\date{March 31, 2015 / version 39}

\begin{abstract}
  
The dark photon, $A^\prime$, and the dark Higgs boson, $h^\prime$, are hypothetical constituents featured in a number of recently proposed Dark Sector Models. Assuming prompt decays of both dark particles, we search for their production in the so-called Higgs-strahlung channel, $e^+e^- \rightarrow A^\prime h'$, with $h^\prime \rightarrow A^\prime A^\prime$. We investigate ten exclusive final-states with $A^\prime \rightarrow e^+e^-$, $\mu^+\mu^-$, or $\pi^+\pi^-$, in the mass ranges $0.1$~GeV/$c^2$~$< m_{A^\prime} < 3.5$~GeV/$c^2$ and $0.2$~GeV/$c^2$~$< m_{h'} < 10.5$~GeV/$c^2$. We also investigate three inclusive final-states, $2(e^+e^-)X$, $2(\mu^+\mu^-)X$, and $(e^+e^-)(\mu^+\mu^-)X$, where $X$ denotes a dark photon candidate detected via missing mass, in the mass ranges $1.1$~GeV/$c^2$~$< m_{A^\prime} < 3.5$~GeV/$c^2$ and $2.2$~GeV/$c^2$~$< m_{h'} < 10.5$~GeV/$c^2$. Using the entire $977\,\mathrm{fb}^{-1}$ data set collected by Belle,  we observe no significant signal.
We obtain individual and combined 90$\%$ credibility level upper limits on the branching fraction times the Born cross section, 
$\cal B \times \sigma_{\mathrm{Born}}$, on the Born cross section, $\sigma_{\mathrm{Born}}$, and on the dark photon coupling
to the dark Higgs boson times the kinetic mixing between the Standard Model photon and the dark photon, $\alpha_D \times \epsilon^2$.
These limits improve upon and cover wider mass ranges than previous experiments. The limits from the final-states $3(\pi^+\pi^-)$ and $2(e^+e^-)X$ are the first placed by any experiment. For $\alpha_D$ equal to 1/137, $m_{h'}<$ 8 GeV/$c^2$, and 
$m_{A^\prime}<$ 1 GeV/$c^2$, we exclude values of the mixing parameter, $\epsilon$, above $\sim 8 \times 10^{-4}$.
    
\end{abstract}


\maketitle

Recent results from dedicated dark-matter searches~\cite{DAMA,CDMS-II,CoGeNT}, muon-spin precession measurements~\cite{g2}, and space-based particle observatories~\cite{PAMELA,Fermi,AMSII} may be interpreted as deviations from the Standard Model (SM) of particle physics. Attempts at devising unified explanations have led to Dark Sector Models (DSM) that introduce a new hidden or dark U(1) interaction that imbues dark matter with a novel charge~\cite{Fayet1,Fayet2,Fayet3,Boehm,Fayet4,Arkani2008,Pospelov2007,Chun2008,Cheung2009,Katz2009,Morrissey2009,Goodsell2009,Baumgart2009,Nomura2008,Alves2009,Jaeckel2010,Batell2009,Reece2009,Essig2009,Bossi2009,Yin2009}. A possible
 mediator of this new Abelian force is the dark photon, which has an expected mass of the order of
MeV/$c^2$~--~GeV/$c^2$ and has a very small kinetic mixing with the Standard Model photon, $\epsilon$, of the order of $10^{-5}$--$10^{-2}$~\cite{Arkani2008}.  
The dark U(1) symmetry group could be spontaneously broken, by a Higgs mechanism, adding a dark Higgs boson $h'$ (or several of these) to such models~\cite{Batell2009}.

Due to the small coupling to SM particles and the low expected mass of the dark photon, the ideal tools to discover the dark photon and the dark Higgs boson are low energy and high-luminosity experiments such as
Belle at KEKB, Belle II at SuperKEKB~\cite{Batell2009}, BaBar at PEP-II~\cite{BaBar,BaBarII}, and dedicated fixed target and beam dump experiments, several of which are
 planned or under construction~\cite{HPS,APEX,DarkLight,BDX,A1_0,A1_1,A1_2}.
This article focuses on
 the Higgs-strahlung channel, $e^+e^- \rightarrow A'h'$.
Generally, the dark photon $A'$ can decay into lepton pairs, hadrons, or invisible particles while the dark Higgs boson
$h'$ can decay into either $A'A'^{(*)}$, leptons pairs, or hadrons, where $A'^*$ is a virtual dark photon. The decay modes of the $A'$ and
 $h'$ depend on their masses and decay 
lengths~\cite{Batell2009,Essig2010}. There are three main cases: (a) $m_{h'} < m_{A'}$: $h'$ is long-lived and decays to lepton pairs or hadrons, (b) $m_{A'} < m_{h'} < 2m_{A'}$: $h' \rightarrow A'A'^*$, where $A'^*$ decays into leptons, and
(c) $m_{h'} > 2m_{A'}$: $h' \rightarrow A'A'$. This article is concerned with case (c); in particular, we investigate ten
exclusive final-states of type $3(l^+l^-)$, $2(\l^+\l^-)(\pi^+\pi^-)$, $2(\pi^+\pi^-)(l^+l^-)$, and $3(\pi^+\pi^-)$, where $l^+l^-$ is an electron or muon pair but not a tau pair, and three inclusive final-states of type $2(l^+l^-)X$, where $X$ is a dark photon candidate detected via missing mass.

The Higgs-strahlung channel involves the effective coupling
 of the dark photon to SM particles, $\alpha'$, induced via kinematic mixing with the SM photon, and the coupling of the dark-photon to the dark Higgs boson, $\alpha_D$. KLOE and BaBar have reported searches for the 
dark photon and the dark Higgs boson~\cite{KLOE,BaBar}: KLOE focused on
 $m_{h'} < m_{A'}$ and BaBar on $m_{h'} > 2m_{A'}$ (assuming prompt decays of the $A'$ and $h'$), but no signal 
was found in either case. BaBar set limits on the product
$\alpha_D \times \epsilon^2$ (where $\epsilon^2 = \alpha'/\alpha_{\rm em}$ and $\alpha_{\rm em}$ is the 
SM electromagnetic coupling constant) for dark photon and dark Higgs boson mass ranges of 0.25~--~3.0~GeV/$c^2$ and 
0.8~--~10.0~GeV/$c^2$, respectively. Beam dump experiments~\cite{Blumlein:1991xh,Blumlein:1990ay,Blumlein:2013cua,Barabash:1992uq,Blumlein2011,E137,E141,E774} have placed 90$\%$~confidence level upper limits on $\epsilon$ for the processes $e^-p \rightarrow A'X'$ and $pp \rightarrow A'X'$ (where $X'$ is not identified) of $\epsilon < 10^{-4}$ for a dark photon mass range of 1~--~300~MeV/$c^2$. Recently, BaBar~\cite{BaBarII} set an upper limit of $\epsilon < 3 \times 10^{-3}$ for a dark photon mass range of 0.3~--~10~GeV/$c^2$ for the radiative decay process $e^+e^- \rightarrow \gamma A'$. The advantage of the Higgs-strahlung
channel compared to the radiative decay is that the Quantum Electrodynamic (QED) background is expected to be much smaller. If, in
addition, the coupling between the dark photon and the dark Higgs boson is of order unity, then the Higgs-strahlung channel is the most sensitive probe for the dark photon.

Here, we report individual upper limits on the branching fraction times the Born cross section, $\cal B \times \sigma_{\mathrm{Born}}$, for the thirteen aforementioned  
Higgs-strahlung final states as well as combined upper limits on $\sigma_{\mathrm{Born}}$ and on the product $\alpha_D \times \epsilon^2$ for these final-states, in the mass ranges $0.1$~GeV/$c^2$~$< m_{A^\prime} < 3.5$~GeV/$c^2$ and $0.2$~GeV/$c^2$~$< m_{h'} < 10.5$~GeV/$c^2$, assuming prompt decays of the dark particles. We use data collected with the Belle detector~\cite{setup} at the KEKB $e^+ e^-$ collider~\cite{kek}, amounting to 977 $\mathrm{fb}^{-1}$ 
at center-of-mass energies corresponding to the $\Upsilon$(1S) to $\Upsilon$(5S) resonances and in the nearby continuum.

\begin{figure}[tht]
\begin{center}
\includegraphics[width=8cm,height=5cm]{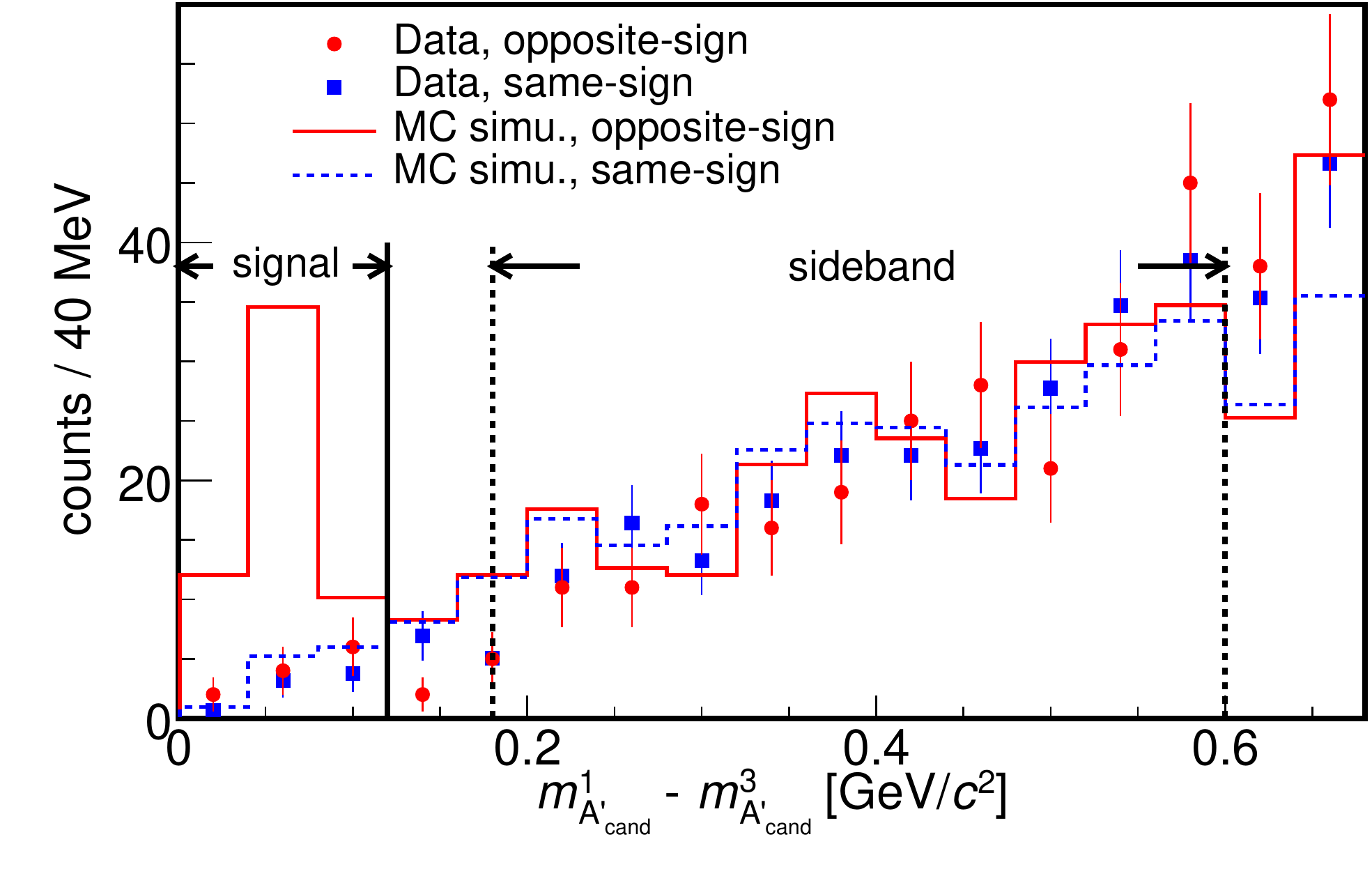}
\caption{Example $m^1_{A'_{\mathrm{cand}}} - m^3_{A'_{\mathrm{cand}}}$ distribution for the $A'h'\rightarrow A'A'A'\rightarrow 6\pi$ channel, for $m^1_{A'_{\mathrm{cand}}}$ = $2.0 \pm 0.1$ GeV/$c^2$, where $m^1_{A'_{\mathrm{cand}}}$ and $m^3_{A'_{\mathrm{cand}}}$ are the dark photon candidates with the highest and lowest mass, respectively. The 
``same-sign'' distributions (blue), where at least one $A'$ candidate is reconstructed from  $\pi^+\pi^+$ or $\pi^-\pi^-$, are normalized to the ``opposite-sign'' 3($\pi^+\pi^-$) distributions (red) in the sidebands, and are used to predict the background in the signal region. 
}
\label{fig1}
\end{center}
\end{figure}

We optimize the selection criteria and determine the $e^+e^- \rightarrow A'h'$  signal detection efficiency using a Monte Carlo (MC) simulation where the interaction kinematics and detector response are simulated with the packages MadGraph~\cite{MadGraph} and GEANT3~\cite{GEANT4}, respectively. There is no suitable background simulation available, so background samples are taken from data sidebands. 

\begin{figure}[tht]
\begin{center}
\includegraphics[width=8cm,height=8cm]{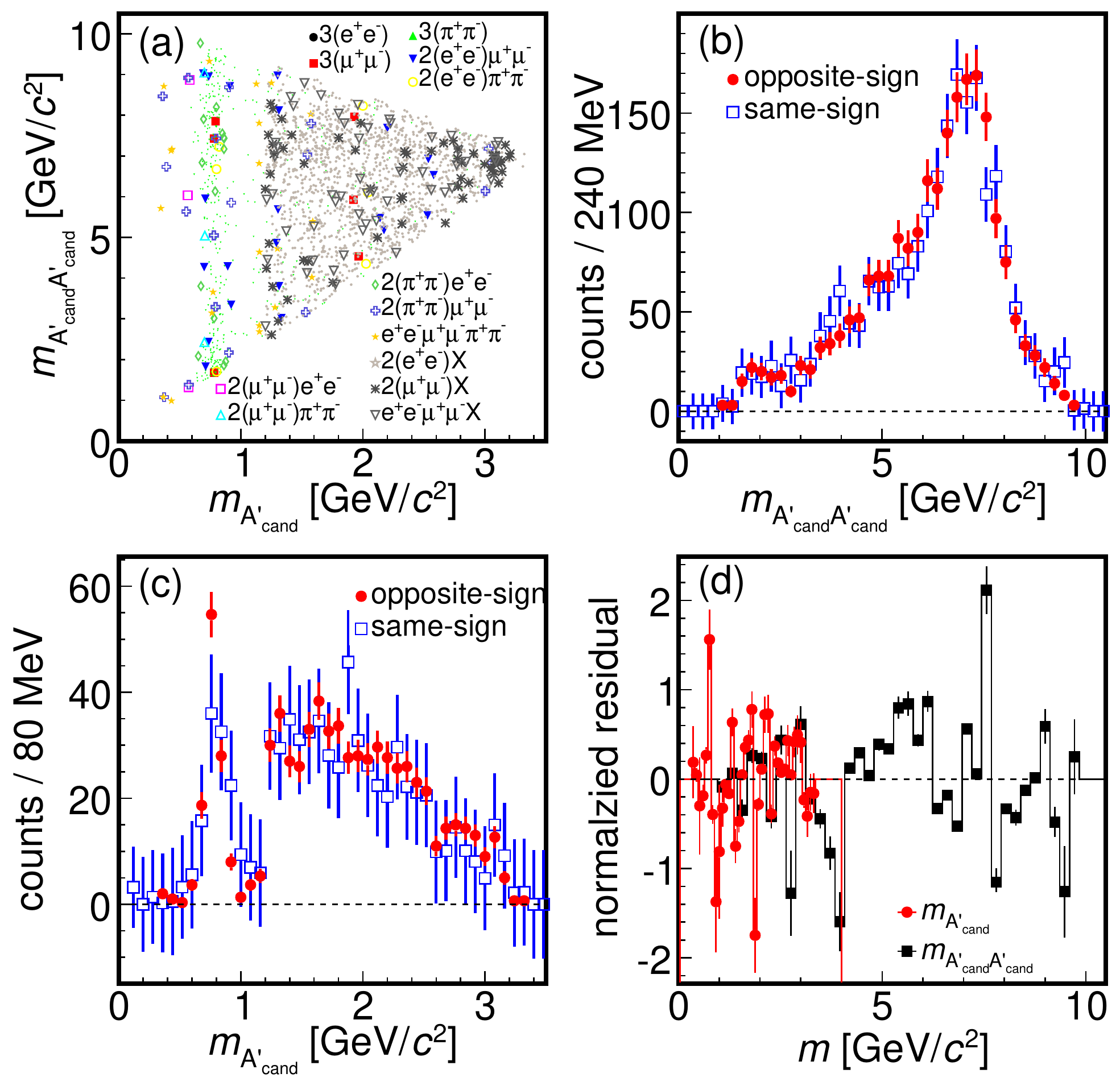}
\caption{(a) : Signal candidates observed versus dark photon candidate mass, $m_{A'_{\mathrm{cand}}}$, and dark Higgs boson
 candidate mass, $m_{A'_{\mathrm{cand}}A'_{\mathrm{cand}}}$, for the 13 final states. There are three entries per event. (b) and (c): Projection of signal candidates onto $m_{A'_{\mathrm{cand}}A'_{\mathrm{cand}}}$ and $m_{A'_{\mathrm{cand}}}$  (red points) with the predicted background (blue squares) from the scaled same-sign distributions for comparison. The dark photon candidate mass distribution has been scaled by 1/3. (d): Normalized 
residuals between the signal candidate distribution and predicted background, versus dark photon candidate mass (red points) and dark Higgs boson candidate mass (black squares). The same-sign error bars
 contain statistical and systematic errors. For empty bins, the systematic error is one event.}
\label{fig2}
\end{center}
\end{figure}

\begin{figure*}[tht]
\begin{center}
\includegraphics[width=12cm,height=7.2cm]{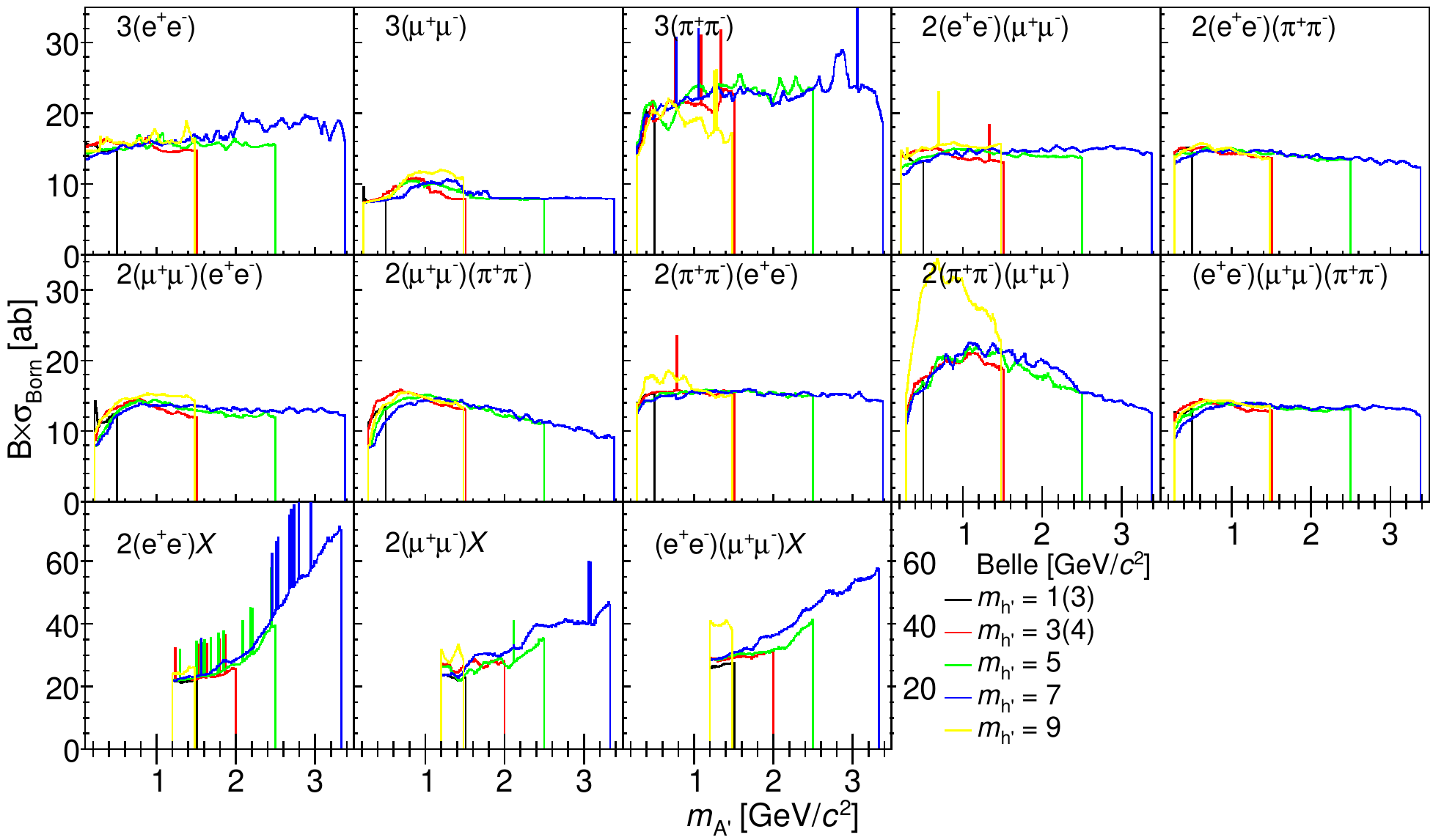}
\includegraphics[width=4cm,height=7.7cm]{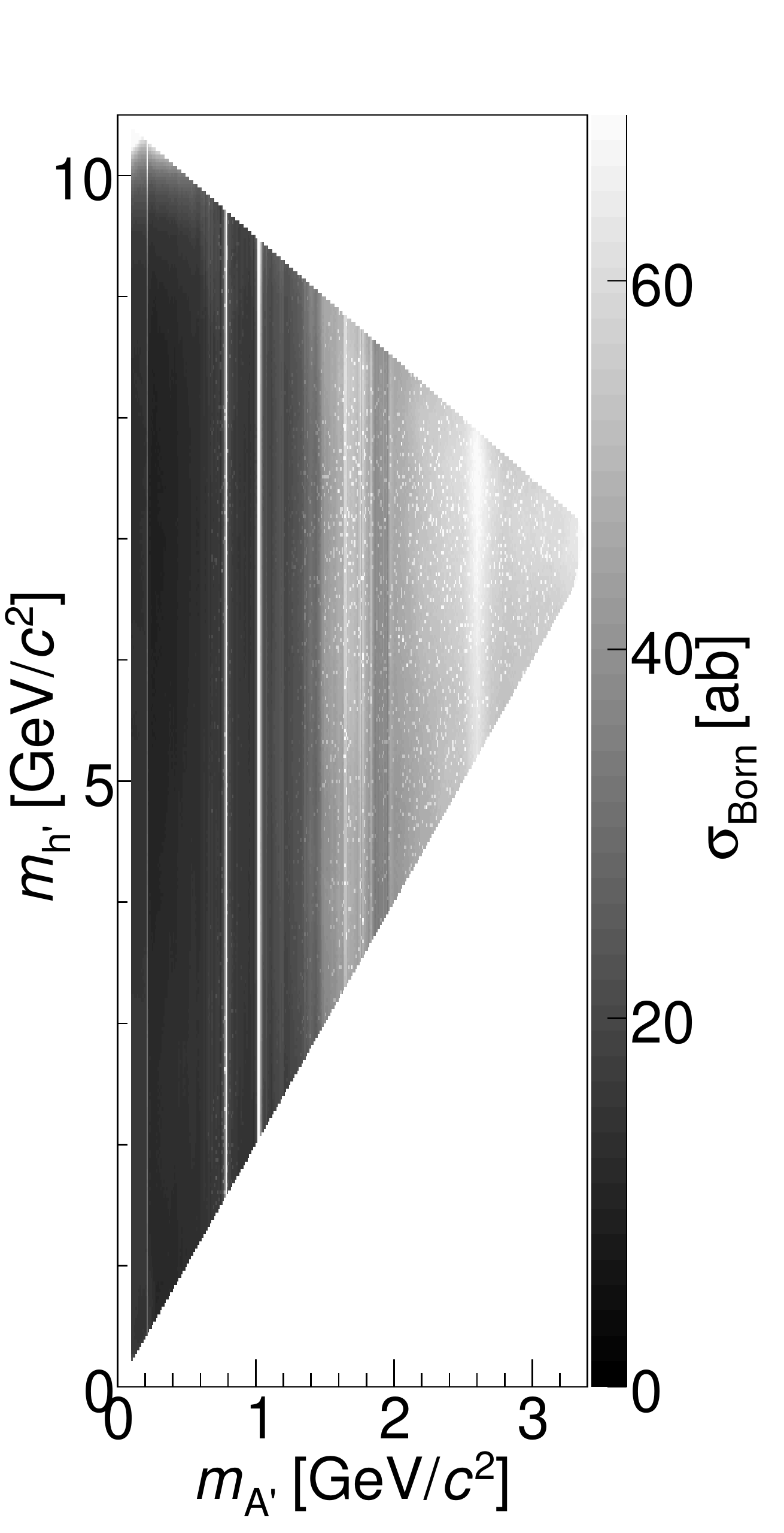}
\caption{Left: 90$\%$ CL upper limit on the product $\cal B \times\sigma_{\mathrm{Born}}$ for each of the 13 final states considered versus dark photon mass for different hypotheses for the dark Higgs boson mass. Black, red, green, blue and yellow curves correspond to $m_{h'}$ = 1, 3, 5, 7, and 9 GeV/$c^2$, respectively, for exclusive channels and $m_{h'}$ = 3, 4, 5, 7, and 9  GeV/$c^2$, respectively, for inclusive channels. Right: 90$\%$ CL upper limit on the cross section of 
$e^+e^- \rightarrow A'h'$, $h' \rightarrow A'A'$ versus dark photon and dark Higgs boson mass.}
\label{fig3}
\end{center}
\end{figure*}
We choose loose particle identification criteria to enhance the detection efficiency of final states with leptons. 
To ensure that only prompt decays are selected, {\it i.e.,} that the decay of each $A'$ candidate occurs near the $e^+ e^-$ interaction point (IP), we require that the
 vertex fit of all tracks detected in the event be consistent with an origin at the IP, and that each track have impact parameters  $|dz| < 1.5$ cm and $dr < 0.2$ cm, where $dz$ is measured along the positron beam (collinear with the $z$ axis) and $dr$ is measured in the transverse $r$--$\phi$ plane. We also require that the second-order Fox-Wolfram moment~\cite{FoxWol} satisfy $R_2$ $<$ 0.9, and that the electron helicity angle, $\alpha_e$, in the $A'$ 
rest frame satisfy $\cos( \alpha_e) < 0.9$, as in Ref.~\cite{BaBar}.   

For exclusive channels, we select final states with exactly three pairs of oppositely charged particles.
For inclusive channels, we select final states of the type
$2(l^+l^-)X$, where $X$ is constrained by the missing mass of the event and contains zero, one, or two reconstructed tracks that are not identified as leptons or pions. We require that both $m_{l^+l^-}$ and $m_X$ be greater than 1.1 GeV/$c^2$. Above this mass, the branching fraction of $A'$ to hadronic final-states other than charged pion pairs is dominant~\cite{Batell2009}. We refer to events selected according to these criteria as ``opposite-sign'' to distinguish them from the ``same-sign'' events used for  background estimation.

For exclusive final states, we select candidate events with final-state masses between $98 \%$ and $ 105 \%$ of the initial-state mass. For inclusive channels, where this condition cannot be applied, we perform a missing-mass analysis: $X$ is treated as an unobserved particle whose missing four-momentum is given by\\
\begin{equation}
 P_{X} = P_{e^+e^-} - P^1_{A'_{\mathrm{cand}}\rightarrow l^+l^-} - P^2_{A'_{\mathrm{cand}}\rightarrow l^+l^-},
\end{equation} where $P_{e^+e^-}$ and $P^{1,2}_{A'_{\mathrm{cand}}\rightarrow l^+l^-}$ are the four-momenta of the initial-state and the two fully reconstructed dark photon candidates, respectively. The mass $m_X$ of the missing four-momentum $P_{X}$ is then compared to the reconstructed
 masses of dark photon candidates $1$ and $2$ using:
\begin{equation}
 \Delta m = m_X - (m^1_{A'_{\mathrm{cand}}\rightarrow l^+l^-} + m^2_{A'_{\mathrm{cand}}\rightarrow l^+l^-})/2.
\end{equation}
We select inclusive final-states by requiring 
\begin{equation}
  \Delta m_{\rm min} < \Delta m < \Delta m_{\rm max},
\end{equation} where the optimized limits $\Delta m_{\rm min}$ and $\Delta m_{\rm max}$ each depend on the measured mean mass of dark photon candidates $1$ and $2$ and on the particular final state.


For exclusive (inclusive) final-states, we then require the invariant masses of dark photon candidates, $m_{A'_{\mathrm{cand}}}$, to be consistent with three (two) distinct  $A' \rightarrow l^+l^-$ or $\pi^+\pi^-$ decays. Signal candidates with three (two) consistent dark photon masses are
 kept by requiring 
\begin{equation}
m_{A'_{\mathrm{cand}}}^{min} < m_{A'_{\mathrm{cand}}} < m_{A'_{\mathrm{cand}}}^{max},
\label{eq2}
\end{equation} where the optimized limits $m_{A'_{\mathrm{cand}}}^{min}$ and $m_{A'_{\mathrm{cand}}}^{max}$ each depend on the measured mean mass of the three (two) fully reconstructed dark photon candidates and on the simulated width of the invariant mass distribution of the dark photon for that mass.

 
For each event, if there is more than one signal candidate that fulfills the selection criteria for a given final state, we select the candidate with the smallest $\Delta m$. For exclusive channels, we use: $\Delta m = \Sigma_1^3 \Delta m_i^2$ with
\begin{equation}
 \Delta m_i = m_{A'_{\mathrm{cand}}}^i - (m_{A'_{\mathrm{cand}}}^1 + m_{A'_{\mathrm{cand}}}^2 + m_{A'_{\mathrm{cand}}}^3)/3. 
\end{equation}
If an event satisfies the selection criteria for multiple final states, we allocate the event to a single final state to ensure that the datasets for each final state are statistically independent. This is accomplished by selecting the lowest numbered final-state category from the following list: (1) exclusive with 
6 leptons, (2) exclusive with four leptons, (3) exclusive with two
 leptons, (4) exclusive with six pions, and (5) inclusive final-states.  
For the signal MC simulation, the fraction of events with multiple signal candidates ranges from $7\%$ to $15\%$ in the channels where we need to apply this ordering. For data, the fraction is below $0.5\%$ in all final states.

We optimize the event selection, including particle identification, the final-state mass requirements, and the parameters $\Delta m_{\rm min}$, $\Delta m_{\rm max}$, $m_{A'_{\mathrm{cand}}}^{min}$ and $m_{A'_{\mathrm{cand}}}^{max}$ using the signal MC simulation only. Events reconstructed as described above are used for signal. Background distributions are derived from the same event sample, by using events where at least
 one dark photon candidate is reconstructed from two tracks with charges of the same sign, enforcing all selection criteria except charge conservation. We refer to these as ``same-sign'' events. We verify that the background 
estimation is consistent with data as shown in Fig.~\ref{fig1}.  We generate MC with specific dark photon and dark Higgs boson masses and interpolate between samples where necessary. 
The detection efficiencies are  20$\%$ and 30$\%$, on average, for the $3(e^+e^-)$ and $3(\mu^+\mu^-)$ final-states, respectively.

For setting limits, we also estimate the background using ``same-sign'' events, but in this case they are from experimental data. 
We sort the dark photon candidates by mass in descending order, $m^1_{A'_{\mathrm{cand}}} > m^2_{A'_{\mathrm{cand}}} > m^3_{A'_{\mathrm{cand}}}$, and calculate the mass
difference $m^1_{A'_{\mathrm{cand}}} - m^3_{A'_{\mathrm{cand}}}$. We divide the data into different bins of $m^1_{A'_{\mathrm{cand}}}$, with each bin
analyzed separately. We divide the $m^1_{A'_{\mathrm{cand}}} - m^3_{A'_{\mathrm{cand}}}$ distribution into two regions: signal and sideband. The signal region size is determined by equation \eqref{eq2}. The sideband region starts at 1.5 times and ends at 5.0 times the signal-region upper limit. Figure~\ref{fig1} 
shows the distribution of the mass difference $m^1_{A'_{\mathrm{cand}}} - m^3_{A'_{\mathrm{cand}}}$ for the bin $m^1_{A'_{\mathrm{cand}}}$ = 2.0 $\pm$ 0.1  GeV/$c^2$ for the six-pion
 final-state.  We assume that, in the absence of signal, the same-sign and the opposite-sign distributions have the same shape (but different yields) in both the signal region and the sideband.  Therefore, for each $m^1_{A'_{\mathrm{cand}}}$ bin, the same-sign 
 distribution (blue squares) is scaled so that the number of events in the sideband agree with the number of opposite-sign events (red points) in the sideband. The expected background in the signal region is then the scaled number 
 of events of the same-sign distribution in that region. This procedure is illustrated by Fig.~\ref{fig1}. The opposite-sign and scaled same-sign distributions are consistent in the signal region and the sideband. In the presence of signal, we would expect an excess of opposite-sign events over the predicted background in the signal region, as can be seen for the signal MC distribution.
\begin{table}
  \caption{Number of events observed after all selection criteria are applied.}
  \begin{tabular}{|c|c||c|c|}
    \hline
    Final-state & Events & Final-state & Events \\
    \hline
    $3(e^-e^+)$ & 1 & $2(\mu^+\mu^-)(e^+e^-)$  & 1 \\
    $3(\mu^+\mu^-)$ & 2 &$2(\mu^+\mu^-)(\pi^+\pi^-)$  & 1  \\
    $3(\pi^+\pi^-)$ & 147 & $2(\pi^+\pi^-)(e^+e^-)$  &  5 \\
    $2(e^+e^-)(\mu^+\mu^-)$ & 7 & $2(\pi^+\pi^-)(\mu^+\mu^-)$ & 6  \\
    $2(e^+e^-)(\pi^+\pi^-)$ & 2 & $(e^+e^-)(\mu^+\mu^-)(\pi^+\pi^-)$  & 7 \\
    $2(e^+e^-)X$ & 572 & $(e^+e^-)(\mu^+\mu^-)X$ & 30 \\
    $2(\mu^+\mu^-)X$ & 20 & & \\
    \hline
  \end{tabular}
  \label{Tab1}
\end{table}
Figure~\ref{fig2} summarizes the background estimation. 
Figure~\ref{fig2} (a) shows the distribution of events measured as a function of the dark photon candidate mass, $m_{A'_{\mathrm{cand}}}$, and the
dark Higgs boson candidate mass, $m_{A'_{\mathrm{cand}}A'_{\mathrm{cand}}}$. Table~\ref{Tab1} shows the number of events observed after all selection criteria are applied.

Figures~\ref{fig2} (b) and (c)
 show the projections on the mass axis of the  dark Higgs boson and dark photon, respectively. 
The number of events observed in the signal region, $N_\mathrm{obs}$, and the number of predicted background events, $N_\mathrm{bkg}$, are in good agreement. Their differences are 
quantified by the normalized residuals, shown in Fig.~\ref{fig2} (d) and defined as 
$(N_\mathrm{obs} - N_\mathrm{bkg}) / \sqrt{\sigma_\mathrm{obs}^2 + \sigma_\mathrm{bkg}^2}$, where $\sigma_\mathrm{obs}$ and $\sigma_\mathrm{bkg}$ are the standard deviations of the distributions. In all cases, the number of events observed is
 consistent with the background estimate. For exclusive final states, the background is mostly due to processes with $\rho$ and $\omega$ resonance particles, such as SM 2$\gamma$ processes. The
discontinuity at 1.1 GeV/$c^2$ in Fig.~\ref{fig2} (c)  is an artifact of the selection criteria. 

\begin{figure*}[tht]
\begin{center}
\includegraphics[width=17cm,height=7.2cm]{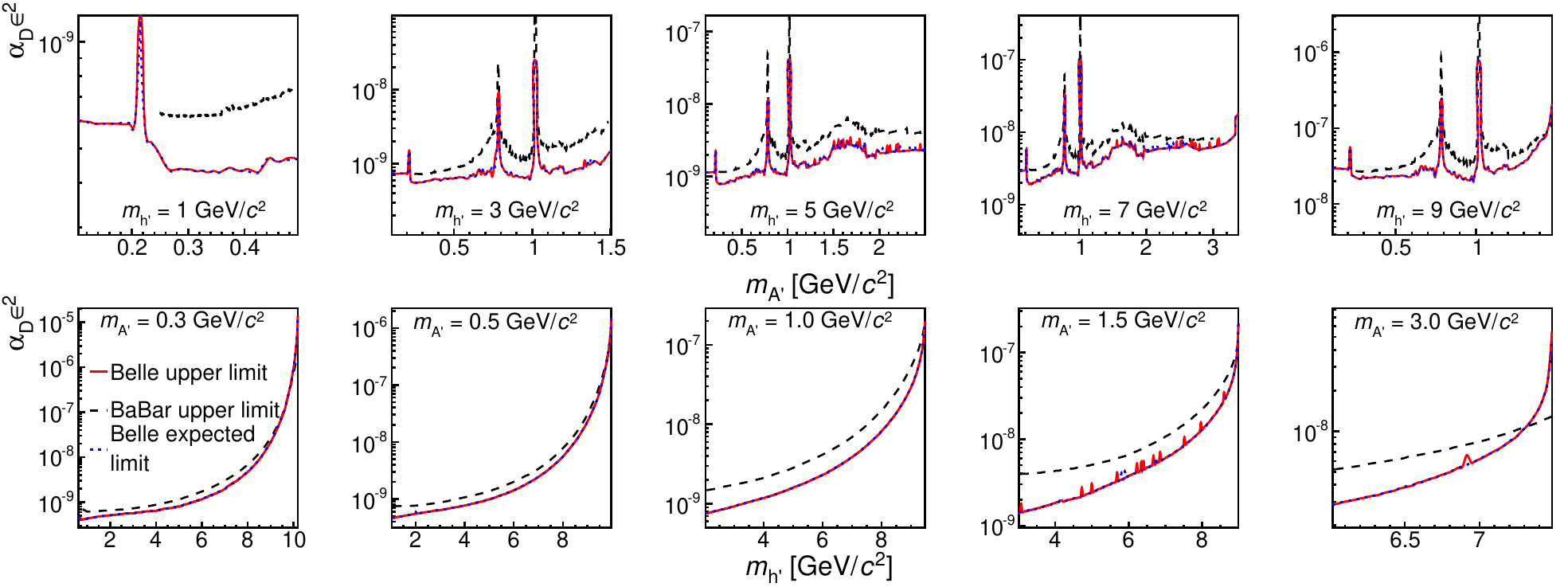}
\caption{90$\%$ CL upper limit on the product $\alpha_D \times \epsilon^2$ versus dark photon 
mass (top row) and dark Higgs boson mass (bottom row) for Belle (solid red curve) and BaBar~\cite{BaBar} 
(dashed black curve). BaBar limits should be divided by ($1+\delta$) before being compared
with Belle limits. The blue dotted curve, which coincides more or less with the solid red curve, shows the expected Belle limit.}
\label{fig4}
\end{center}
\end{figure*}

The upper limits on $\cal B \times \sigma_{\mathrm{Born}}$ and $\sigma_{\mathrm{Born}}$ are calculated for ranges of $m_{A'}$ and $m_{h'}$, based on the signal MC mass resolution, with a Bayesian inference method with the use of Markov Chain Monte Carlo~\cite{BAT}. The 
number of observed events can be expressed as:
\begin{equation}
 N_{\mathrm{obs}} = \sigma_{\mathrm{Born}} \cdot (1 + \delta) \cdot |1-\Pi|^2 \cdot {\cal L}  \cdot {\cal B} \cdot \varepsilon + N_{\mathrm{bkg}},
\end{equation} where $1 + \delta$ is an initial-state radiative correction factor, $|1-\Pi|^2$ is the vacuum polarization factor, $\cal L$ is the luminosity, $\varepsilon$ is the detection efficiency, and $N_{\mathrm{bkg}}$ is the number of predicted background events. We calculate, for the exclusive (inclusive) channels, $1 + \delta$ using the formulae in Ref. \cite{ISRa} and assuming the theoretical cross section is proportional to $1/s$~\cite{Batell2009}, where $s$ is the square of the initial-state mass, and also assuming a cut-off value corresponding to $98 \%$ (a value between $20 \%$ and $90 \%$) of the initial-state mass. $1+\delta$ varies from 0.804 (0.93) to 0.807 (1.17) depending on $s$ and for the inclusive channels also the effective cut-off value.  We use $1+\delta$ = 0.8055 (1.0) and include the variation as a systematic error in the upper limit calculation. The value of $|1-\Pi(s)|^2$ is taken from Ref.~\cite{ISRb,Fedov} and varies between 0.9248 and 1.072 depending on $s$. For 
$\cal B \times \sigma_{\mathrm{Born}}$ and $\sigma_{\mathrm{Born}}$, logarithmic priors are used, and for $1 + \delta$, $|1-\Pi|^2$, ${\cal L}$, ${\cal B}$, $\varepsilon$, and $N_{\mathrm{bkg}}$
 Gaussian priors are used to take into account the systematic uncertainty. 
In Fig.~\ref{fig3}, the left panel shows the 90$\%$ credibility level (CL)~\footnote[1]{In common High Energy Physics usage, this credibility level has been reported as ``confidence level'', which is a frequentist-statistics term.} upper limits  on $\cal B \times \sigma_{\mathrm{Born}}$ versus the dark photon mass, for different
 hypotheses of the dark Higgs boson mass, for each of the 13 final states considered, while the right panel shows the combined upper limit on $\sigma_{\mathrm{Born}}$ for $e^+e^- \rightarrow Ah'$ versus the dark photon and dark Higgs boson mass. For the combined limit, compared to BaBar, we use two extra
 channels, $3(\pi^+\pi^-)$ and $2(e^+e^-)X$, which contribute 91$\%$ of our background. The branching fractions were taken from Ref.~\cite{Batell2009}.

The combined limit can also be expressed as a limit on the product $\alpha_D \times \epsilon^2$ by using the equations described in Ref.~\cite{Batell2009}. Figure~\ref{fig4} shows the 90$\%$ CL upper limits on $\alpha_D \times \epsilon^2$ for Belle, expected and measured, and for BaBar, for five different mass hypotheses for the dark Higgs boson (top row) and dark photon (bottom row) masses. Note that the BaBar limits were based on the visible cross section, rather than the Born cross section. For the expected limit, we assume: $N_{\mathrm{obs}} = N_{\mathrm{bkg}}$.

The inclusion of the $3(\pi^+\pi^-)$ final state dramatically improves the limit around the $\rho$ and $\omega$ resonances.  
 The dominant sources
of systematic uncertainties are: the integrated luminosity (1$\%$), branching fractions (4$\%$), track identification (6$\%$), particle identification efficiency (5$\%$), detection efficiency (15$\%$),
background estimation (10$\%$) and initial-state radiation (15$\%$). All systematic uncertainties added in quadrature amount to 25$\%$.  

In summary, we search for the dark photon and the dark Higgs boson in the mass ranges 0.1~--~3.5~GeV/$c^2$ and 0.2~--~10.5~GeV/$c^2$, respectively. No significant signal is observed. We obtain individual and combined 90$\%$ CL upper limits
 on the product of branching fraction times the Born cross section, $\cal B \times \sigma_{\mathrm{Born}}$, on the Born cross section, $\sigma_{\mathrm{Born}}$, and on the product of the dark
 photon coupling to the dark Higgs boson and the kinetic mixing between the Standard Model photon and the dark photon, $\alpha_D \times \epsilon^2$. 
These limits improve upon and cover wider mass ranges than previous experiments and the limits in the final-states $3(\pi^+\pi^-)$ and $2(e^+e^-)X$, 
where $X$ is a dark photon candidate detected via missing mass, are the first limits placed by any experiment. For $\alpha_D$ equal to 1/137, $m_{h'}<$ 8 GeV/$c^2$, and 
$m_{A'}<$ 1 GeV/$c^2$, we exclude values of the mixing parameter, $\epsilon$, above $\sim 8 \times 10^{-4}$. In the mass ranges, and for modes, where previous measurements from BaBar exist, the limits reported here are almost a factor of two smaller. Since the backgrounds are very low to non-existent, the improvement scales nearly linearly with the integrated luminosity.
 This bodes well for future searches with Belle II.

\begin{acknowledgements}
We thank the KEKB group for excellent operation of the
accelerator; the KEK cryogenics group for efficient solenoid
operations; and the KEK computer group, the NII, and 
PNNL/EMSL for valuable computing and SINET4 network support.  
We acknowledge support from MEXT, JSPS and Nagoya's TLPRC (Japan);
ARC (Australia); FWF (Austria); NSFC (China); MSMT (Czechia);
CZF, DFG, and VS (Germany); DST (India); INFN (Italy); 
MOE, MSIP, NRF, GSDC of KISTI, and BK21Plus (Korea);
MNiSW and NCN (Poland); MES and RFAAE (Russia); ARRS (Slovenia);
IKERBASQUE and UPV/EHU (Spain); 
SNSF (Switzerland); NSC and MOE (Taiwan); and DOE and NSF (USA).
We would also like to thank Bertrand Echenard for discussing the BaBar analysis and Rouven Essig for providing the models for MadGraph. 
\end{acknowledgements}


\end{document}